\documentclass[11pt,preprint,flushrt]{aastex}

\newcommand\etal{{\it et al.\/}}
\newcommand\bfa{\mbox{{\boldmath $\alpha$}}}
\newcommand\bfha{\mbox{{\boldmath $\hat\alpha$}}}
\newcommand\bft{\mbox{{\boldmath $\theta$}}}

\begin{document}
 
\slugcomment{Revision 2.0 8/8/00, accepted to AJ}
 
\title{Orbit Fitting and Uncertainties for Kuiper Belt Objects}
\author{Gary Bernstein}
\affil{Department of Astronomy, University of
 Michigan, Ann Arbor, MI 48109-1090}
\email{garyb@astro.lsa.umich.edu}
\and
\author{Bharat Khushalani}
\affil{Department of Aerospace Engineering, University of
 Michigan, Ann Arbor, MI 48109-2140}
\email{Email: bharat@engin.umich.edu} 

\begin{abstract} 
We present a procedure for determination of positions and orbital
elements, and associated uncertainties, of outer Solar System planets.
The orbit-fitting procedure is greatly streamlined compared to
traditional methods because acceleration can be treated as a
perturbation to the inertial motion of the body.
These techniques are immediately applicable to Kuiper Belt Objects,
for which recovery
observations are costly.  Our methods produce positional
estimates and uncertainty ellipses even in the face of the substantial
degeneracies of short-arc orbit fits; the sole {\it a priori}
assumption is that the orbit should be bound or nearly so.  
We use these orbit-fitting
techniques to derive a strategy for determining Kuiper Belt orbits
with a minimal number of observations.
\end{abstract}


\section{Introduction}
The discovery of the Kuiper Belt \citep{Je93} created a new vista upon
the formation and early evolution of the Solar System.  Study of these
objects' basic physical properties 
of course requires sufficiently well-determined orbital parameters
to deduce a future position and the distance to the object.
The orbital parameters are themselves of great interest, with the
phase space of orbits threaded by an intricate web of resonant and
long-lived chaotic zones \citep{Ma96}.  For both the practical goal
of object retrieval and the higher goal of understanding the dynamical
structure of the population, it is essential to not only have accurate
orbits but also to quantify the {\em uncertainty} in the orbit.  

There are centuries-old methods for determination of orbital
parameters from observed positions.  Present-day desktop
computers execute these solutions trivially, and a linearized
propagation of the positional uncertainties into the orbital elements
is straightforward as well \citep{Mu93}.  Workstations are even fast
enough to bound the uncertainty region with a brute-force sampling of
the 6-dimensional orbit space in many cases.
Why then should we bother developing
another orbit-fitting technique? KBOs pose a particular challenge
because the objects are faint, so recovery observations are quite
costly, requiring substantial investment of 2--4-meter telescope time
for all but the brightest known objects.  As a consequence, most known
objects have been observed only a few times, leading to substantial
degeneracies in the orbit fits.  We need to know the error ellipse (in
sky position and in orbital parameter space) even in the face of these
degeneracies.  In the best case a brute-force method will work on
underdetermined orbits but leave us ignorant of the nature of the
degeneracy.  In the worst case the degeneracies will lead to long or
effectively infinite computation times.

The Minor Planet Center (MPC)
(http://cfa-www.harvard.edu/iau/mpc.html)
provides orbital elements
and predicted ephemerides for new objects.  The approach of the MPC to
short-arc objects is to select the simplest ``sensible'' orbit that
fits the data---{\it e.g.\/} a circular orbit, if it fits the data and
does not imply a close approach to Neptune; a 3:2 resonant orbit with
Neptune if the circular orbit would imply a close approach; or,
if the circular orbit is a poor fit, a
``V\"ais\"al\"a'' orbit, with the object near perihelion \citep{Ma85,
Ma91}, if this does not imply a close encounter.
The MPC orbits in effect remove the
degeneracies by assuming that the orbit is most likely to resemble
those of known objects, and consequently predict positions that
in most cases are quite close to the actual recovery positions.  There
are a few difficulties with this approach, however:  first, focussing
recovery efforts on the positions predicted by these favored orbits
will inevitably bias the recovered population toward such orbits.
This makes it difficult to quantify the portion of the population
which may be in unusual orbits as they are preferentially lost.
Second, uncertainties on positions and orbital elements
are not currently available from the Minor Planet Center.
Positional uncertainties are essential for proper planning and
evaluation of recovery observations.  Uncertainties on orbital
elements are important for analyses of the dynamical characteristics
of the population.  For example, there was speculation that the 2:1
resonance with Neptune was empty or underpopulated, but 1997 SZ$_{10}$ and
1996 TR$_{66}$ were reclassified from 3:2 resonators to 2:1 resonators
after further observations.  Finally, it is sometimes necessary to
calculate orbits ``on the fly'' during an observing run, and the MPC
should not be expected to provide instantaneous analyses.

\section{Methods}
\subsection{Motivation}
Our approach to orbit-fitting for KBOs is motivated and guided by the
following differences from the historically more common application to
the asteroid population:
\begin{itemize}
\item Recovery observations are very expensive, as noted.  This leads
us to investigate, in \S\ref{optimal}, the
minimal number of observations required to reach a given accuracy in
orbital elements.  The prediction of future positions should be a
stable procedure even in the presence of nearly-degenerate orbital
elements. This will require that we fully understand the nature of the
degeneracies from short arcs.  We would like to find a
parameterization of the orbits in which the degeneracies are confined
to as few parameters as possible.
\item The release of the USNO-A2.0 astrometric catalog\footnote{
The USNO-A2.0 catalog, produced by D. Monet \etal, is described at
http://www.nofs.navy.mil/projects/pmm/USNOSA2doc.html.}
now makes it possible to measure astrometric positions to 0\farcs2
accuracy over the full sky \citep{De99} in a typical CCD image, and
to produce reasonable estimates of the uncertainties on each position
measurement.  Uncertainty estimation can therefore proceed by the
straightforward propagation of measurement errors.
Space telescopes and adaptive-optics ground-based telescopes will
commonly produce relative positions to 0\farcs01 accuracy in the near
future.  Very accurate positions should in principle produce accurate
orbits even over relatively short arcs.
\item KBOs are at distances $d\gtrsim30$~AU so their apparent motions
are dominated by reflex motion.  The observed arcs are a small
fraction of the orbital period $P$ even after a decade.  Apparent
motion patterns are thus very simple and unambiguous in comparison,
say, to Near-Earth Objects.
\item If we parameterize the distance to the object by
$\gamma=(1{\rm\,AU})/d\lesssim0.03$, the acceleration of the KBO is
$\gamma^2\lesssim10^{-3}$ times smaller than Earth's acceleration, and
the component transverse to the line of sight is
$\gamma^3\approx10^{-4.5}$ times Earth's.  The KBO motion is nearly
inertial, and gravitational acceleration can be treated as a
perturbation.  Instead of parameterizing orbits by the usual element
vector ${\bf a}=\{a,e,i,\Omega,\omega,T_p\}$, an orbit is more stably
specified by some Cartesian phase space vector 
${\bf P}=\{x_0,y_0,z_0,\dot x_0, \dot y_0, \dot z_0\}$ 
at the time of initial observation.
\item The annual parallax is limited to $2\gamma\approx3\arcdeg$, and
the total object motion is only $\approx10\arcdeg$ over a decade. We will
therefore express sky positions as coordinates $(\theta_x, \theta_y)$
in a tangent-plane projection of the celestial sphere about an
appropriate reference point, {\it e.g.\/} the first observed
location.  Furthermore, the line of sight to the object changes by
only a few degrees, and we will find it convenient to align our
Cartesian coordinates with the $z$ axis along the initial line of sight.
\item Daily parallax is $\le8\times10^{-5}\gamma\lesssim0\farcs5$.
For most ground-based observations this will be too small to yield
useful information.
\end{itemize}

\subsection{Exact Equations}
The input data are the observed right ascension and declination
$(\alpha_i,\delta_i)$ of the target at times $t_i$.  We will assume
that all positions have been measured in the J2000 celestial reference
frame via astrometry tied to background stars in the USNO-A2 catalog.
Measuring relative to the background stars exactly cancels the effects
of stellar aberration, 
and reduces the maximal error from gravitational
deflection of starlight to the milli-arcsecond level, so we can treat
the problem with Euclidean geometry.
We transform the observed coordinates to a tangent-plane projection
$(\theta_x,\theta_y)$
about a reference location $(\alpha_0, \delta_0)$.  The $\theta_x$
axis is taken in the Eastward direction of the J2000 ecliptic
coordinate system, and $\theta_y$ points toward J2000 ecliptic North.
Formulae for the conversion from $(\alpha,\delta)$ to
$(\theta_x,\theta_y)$ are straightforward and summarized in
Appendix~\ref{sphere}.  Figure~\ref{coords} illustrates the relation
of our coordinate systems to the ecliptic.

\begin{figure}[t]
\plotone{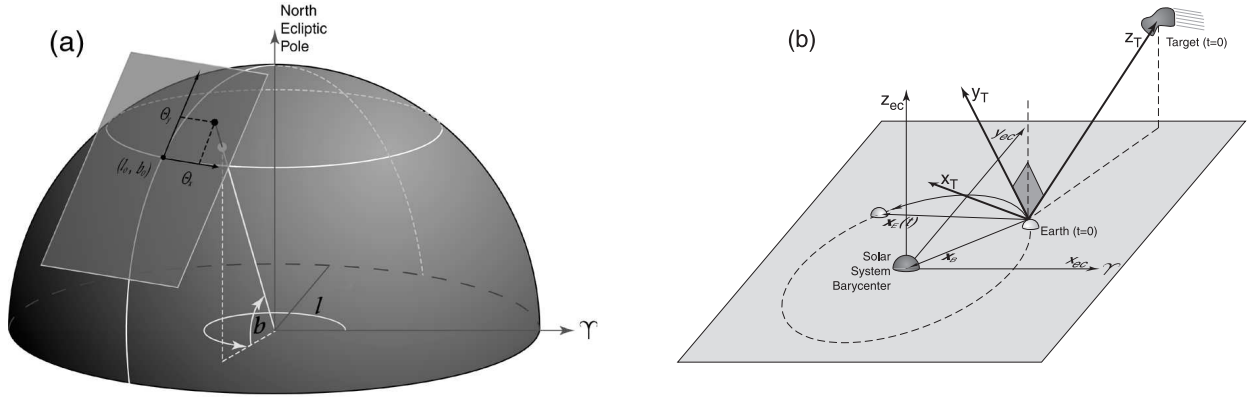}
\caption[dummy]{\small
Panel (a) illustrates the relation of our angular coordinate system to
ecliptic coordinates.  Our coordinates are a tangent-plane projection
of the sky about the initial line of sight, $(\ell_0,b_0)$.  Panel (b)
illustrates the spatial coordinate system ${\bf x}_T$ we adopt in
relation to the ecliptic spatial system ${\bf x}_{\rm ec}$.  Our
system has origin at the observer location on Earth's surface for
$t=0$ and the $\hat z$ axis along the initial line of sight to the
target.  The vectors ${\bf x}_E$ and ${\bf x}_B$ toward the observer
and solar system barycenter are illustrated as well.  Both spatial and
angular $\hat x$ and $\hat y$ axes are aligned to local ecliptic East
and North, respectively.
}
\label{coords}
\end{figure}

We will generally take this projection axis to be the first observed
position, and will take the first observation to be at time $t_0=0$.
We will assume that $\theta_x$ and $\theta_y$ have equal and
uncorrelated measurement uncertainties $\sigma_i$ (generalization to
non-circular error regions is straightforward).

We define our inertial Cartesian coordinate system by placing the
origin at (or near) the location of the observer at the initial epoch
$t=0$.  The $z$ axis is directed toward the reference direction
$(\alpha_0,\delta_0)$ of the tangent-plane projection, {\it i.e.\/}
along the initial line of sight.  The $y$ axis is in the plane
containing the $z$ axis and the J2000 North ecliptic pole, {\it
i.e.\/} parallel to the $\theta_y$ axis of the tangent plane.  The $x$
axis then lies in the ecliptic plane and
points along $\theta_x$ to ecliptic East.  Unless otherwise
specified, we will take the units of distance and time to be AU and
Julian year, respectively.  The coordinate systems are illustrated in
Figure~\ref{coords}. 

If ${\bf x}(t)$ represents the position of the target at time $t$ and
${\bf x}_E(t)$ is the position of the observer on Earth, then the
observed sky position at time $t$ is 
\begin{equation}
\begin{array}{ccc}
\theta_x(t) & = & { {x(t-\Delta t) - x_E(t)} \over {z(t-\Delta t) -
z_E(t)}} \\
\theta_y(t) & = & { {y(t-\Delta t) - y_E(t)} \over {z(t-\Delta t) -
z_E(t)} }
\end{array}
\label{exact}
\end{equation}
where $\Delta t$ is the light-travel time $|{\bf x}(t-\Delta t)-{\bf
x}_E(t)|/c$. The definition of $\Delta t$ is recursive but in practice
is easily solved for iteratively because of the lack of ambiguity in
KBO orbit solutions.

We wish our uncertainties to be limited by the errors in the
astrometric catalog, so our requirement is
to model the target position to $<0\farcs1$ accuracy
over $\approx10$-year time spans.
For target distance $\gtrsim30$~AU, the object and observer positions
must be correct to $\approx2000$~km.
The Earth's position relative to the Sun and Solar System barycenter
(SSB) is available from the JPL DE405 Ephemeris\footnote{
The DE405 ephemeris is available at http://ssd.jpl.nasa.gov/eph\_info.html}
\citep{DE405}
to an accuracy many orders of magnitude better than we require.  A
simplistic model of the Earth as an oblate spheroid in constant
rotation gives the observatory position to accuracy well within our
tolerance, so we may consider ${\bf x}_E$ to be known exactly.
The interpolated JPL ephemeris and topocentric correction are rapidly
calculable. 

\subsubsection{Orbit Approximations}
We write the target orbit as
\begin{equation}
\label{orbit1}
{\bf x}(t) = {\bf x}_0 + {\bf \dot x}_0 t + {\bf g}(t),
\end{equation}
where ${\bf g}$ is the gravitational perturbation satisfying
\begin{eqnarray}
{\bf g}(t=0) & = & \dot {\bf g}(t=0)  =  0, \\
\ddot {\bf g}(t) & = & \sum_j { -{GM_j[{\bf x}(t)-{\bf x}_j(t)]}
	\over {|{\bf x}(t)-{\bf x}_j(t)|^3 } },
\label{nbody}
\end{eqnarray}
where the sum runs over the other bodies in the Solar System.  
We include forces from the Sun and the giant planets, obtaining their
positions from the DE405 ephemeris.  For objects beyond 30~AU, 
one could
approximate the total gravitational field of
the Solar System as a monopole originating
the Solar System barycenter, with errors of
only $\sim10^{-4}$~AU over 10 years (barring Neptune encounters),
yielding positions accurate to $<1\arcsec$.
The computational
expense of the full N-body form Eq.~(\ref{nbody}) is insignificant,
however, as the time steps in the orbit integration can be quite long
(20 days or more) without significant loss of accuracy over a decade.

The combination of Eqs. (\ref{exact}), (\ref{orbit1}), and (\ref{nbody})
define our model for the position $(\theta_x,\theta_y)$ as a function
of time and the initial state vector ${\bf P}$. A comparison of our
positions for Pluto, Neptune, and 1992 QB1 to those generated by the JPL
ephemeris indicate residual differences of $<0\farcs05$ for a 10-year
integration.

\subsection{Approximate Formulae \& Degeneracies}
We can understand the nature of the short-arc degeneracies by
expanding Eq. (\ref{exact}) in powers of time $t$ and distance
parameter $\gamma$.  The formulae are simpler if we redefine our phase
space basis in terms of the vector $\bfa
=\{\alpha,\beta,\gamma,\dot\alpha, \dot\beta, \dot\gamma\}$,
where
\begin{equation}
\label{abgdef}
\begin{array}{lll}
\alpha \equiv {x_0}/{z_0} & \beta \equiv {y_0}/{z_0} & 
	\gamma \equiv 1/{z_0} \\ 
\dot\alpha \equiv {\dot x_0}/{z_0} & \dot\beta \equiv {\dot y_0}/{z_0} & 
	\dot\gamma \equiv {\dot z_0}/{z_0} .
\end{array}
\end{equation}
Please note that $\dot\alpha$, $\dot\beta$, and $\dot\gamma$ quantify
the motion along our three axes but are {\it
not} the time derivatives of $\alpha$, $\beta$, and $\gamma$.
We have chosen our coordinate system so that $\alpha, \beta\approx 0$,
and we expect $\dot\alpha, \dot\beta, \dot\gamma \sim
2\pi\gamma^{3/2}$ for bound orbits.  If the orbit is nearly circular,
the line-of-sight velocity is down another factor of $\gamma$ so that
$\dot\gamma\sim2\pi\gamma^{5/2}$. 
The $x$ component of the apparent motion is now
\begin{eqnarray}
\label{abgposn}
\theta_x & = & { { \alpha + \dot\alpha t + \gamma g_x(t) - \gamma x_E(t) }
 \over { 1 + \dot\gamma t + \gamma g_z(t) - \gamma z_E(t) } } \\
 & \approx & \alpha + (\dot\alpha - \gamma\dot x_E) t \\
 & & 	+ \gamma(\ddot g_x - \ddot x_E) t^2/2 
	- (\dot\gamma - \gamma\dot z_E)(\dot\alpha - \gamma\dot x_E)
 t^2 \nonumber \\
 & & 	+ O(t^3). \nonumber
\end{eqnarray}
In the second line we have made the approximation that $t\ll 1$~year.
The equation for $\theta_y$ is analogous.  We see that the solution
for the orbital parameters $\bfa$ has three regimes:

\subsubsection{Slope Regime}
For arcs spanning consecutive nights, we can only determine the 
instantaneous positions $\alpha$ and $\beta$, plus the slopes
\begin{eqnarray}
\label{slopex}
\dot\theta_x & = & (\dot\alpha - \gamma\dot x_E)
	\approx \dot\alpha - \gamma (2\pi\cos\phi_0) \\
\label{slopey}
\dot\theta_y & = & (\dot\beta - \gamma\dot y_E)
	\approx \dot\beta
\end{eqnarray}
The approximations assume that we are observing near the ecliptic
plane, and that the Earth's orbit is circular with phase $\phi_0$ at
time $t=0$ relative to the Sun-target vector.  In this situation the
line-of-sight motion $\dot\gamma$ is completely unconstrained, and
there is a total degeneracy between the distance $\gamma$ and the
ecliptic motion $\dot\alpha$.  For a circular orbit,
$|\dot\alpha|\approx (\sqrt{\gamma}/\cos\phi_0) (\gamma\dot x_E)$, and
for any bound orbit $|\dot\alpha|< (\sqrt{2\gamma}/\cos\phi_0)
(\gamma\dot x_E)$.  The usual observational strategy, therefore, is to
observe near opposition ($\phi_0=0$) to minimize the proper-motion
term $\dot\alpha$ relative to the parallax term $\gamma\dot x_E$, 
giving an estimate of $\gamma$ to fractional
accuracy of $\sqrt{2\gamma}\approx0.25$ or better.  An ephemeris may
be predicted under the assumption of a prograde
circular orbit.  Below we will
describe a simple way to place an error ellipse on the position without
assuming a circular orbit.

\subsubsection{Acceleration Regime}
As the arc length grows, the next bit of information gleaned is the
apparent acceleration vector:
\begin{eqnarray}
\label{accelx}
\ddot\theta_x & = & 
	\gamma(\ddot g_x - \ddot x_E) 
	-2 (\dot\gamma - \gamma\dot z_E)(\dot\alpha - \gamma\dot x_E)
	\\
& \approx & \gamma(\ddot g_x - 4\pi^2\sin{\phi_0}) 
	+ 2 \dot\gamma\ (2\pi\gamma\cos{\phi_0}).
\end{eqnarray}
The second line again assumes an ecliptic observation and a circular
Earth orbit, and keeps leading-order terms only.  We see first that
the effect of $\dot\gamma$ on the acceleration is a factor
$\gamma/2\pi<0.005$ smaller than the $\gamma\ddot x_E$ term.  In this
regime, therefore, $\dot\gamma$ is still indeterminate.  The
target's transverse acceleration $\ddot g_x$ is smaller by a factor
$\gamma^3$ than the Earth's acceleration and can also be ignored at
this point.  The apparent acceleration is then
\begin{eqnarray}
\label{accel}
\ddot\theta_x & \approx &
-\gamma(4\pi^2\sin{\phi_0})
\\
\ddot\theta_y & \approx & 0
\nonumber
\end{eqnarray}
We see that the apparent acceleration is a robust and simple way to
constrain $\gamma$, the distance to the target.  This method fails
however, for observations at opposition ($\phi_0=0$), where the reflex
acceleration vanishes.  

What time span is needed to measure the acceleration to a useful
accuracy?  Let us presume that the object is initially detected in a
pair of observations separated by 24 hours, then recovered in a single
observation some time $T$ later, and that each observation has
astrometric uncertainty $\sigma$.  Using Eq. (\ref{accel}) to
determine $\gamma$, the fractional error in the
distance to the target is then roughly 
\begin{eqnarray}
\label{disterr}
{\sigma_d \over d} & = & {\sigma_\gamma \over \gamma} 
\approx { \sigma \over {\sqrt{2}\pi^2 \gamma \sin{\phi_0} (24\,{\rm
h}) T} } \\
 & = & 7.5\% \left({\sigma \over 0\farcs2}\right)
	\left({ 1{\,\rm week} \over T}\right)
	\left({\sin 45\arcdeg \over \sin \phi_0}\right)
	\left({d \over 40{\,\rm AU}}\right)  \nonumber
\end{eqnarray}
The distance to most KBOs can be determined from the apparent
acceleration with only 1 week's arc,
{\it as long as the observations are not near opposition.}  
For a main belt asteroid, the apparent acceleration is an order of
magnitude larger and is easily detected in a 24-hour span.  This means
it is easy to distinguish main belt asteroids from KBOs away from
opposition, even near the main-belt turnaround points.

A typical (non-opposition) 1-week arc thus gives a useful estimate of
5 of the 6 phase-space parameters of the orbit, with the line-of-sight motion
being completely indeterminate.  In this basis the degeneracy is
confined to a single parameter, whereas the degeneracy is shared
between most of the traditional orbital elements.  In particular
neither $a$ nor $e$ is at all well constrained at this point.

It is interesting to note that the orbital angular momentum is
sensibly constrained in this regime, since the line-of-sight
velocity is nearly radial to the Sun and contributes little to the
angular momentum.  The energy of the orbit, however, is
\begin{equation}
\label{energy}
E = -GM_\odot\gamma (1 - 2\gamma\cos\beta_0 + \gamma^2)^{-1/2}
	+ (\dot\alpha^2 + \dot\beta^2 + \dot\gamma^2)/2\gamma^2,
\end{equation}
and is uncertain until $\dot\gamma$ can be constrained.  Here
$\beta_0$ is the solar elongation of the target at the initial epoch.

\subsubsection{Fully Constrained Regime}

The line-of-sight velocity can be constrained as the Earth moves
around its orbit to give us a new point of view.  From Eq. (\ref{abgposn}),
the leading $\dot\gamma$ term in the sky position is $\dot\gamma t
\gamma x_E(t) \lesssim \dot\gamma\gamma t$.  We know that
$\dot\gamma^2<2GM_\odot \gamma^3$ from the simple constraint that the
object be bound.  A useful constraint on $\dot\gamma$ is obtained only
after sufficient time that a $\dot\gamma$ value below the
binding limit would produce a measured displacement that is
distinguishable from terms in the other 5 parameters.  This can occur
on a timescale of months for high-quality observations.  The
uncertainty in $\dot\gamma$ (or $\dot z$) dominates the phase space
uncertainties for essentially all lengths of arc, though it does
not always dominate the ephemeris uncertainties.

\subsection{Fitting Procedures}
With 3 or more observations spanning several months or more, we will
be in the fully-constrained regime.  At this point the constraint of
the orbital parameter vector $\bfa$ from the measured
positions $\bft_i$ becomes a straightforward
non-linear minimization problem.  The formalism for Bayesian orbit
fitting is discussed in detail by \citet{Mu93}.  We will assume
Gaussian measurement
errors, in which case the most likely parameter vector $\bfha$ is 
the least-squares value, minimizing
\begin{equation}
\label{chisq}
\chi^2 = \sum_i \left\{ 
	{[\theta_{x,i} - \theta_x(\bfha,t_i)]^2 \over \sigma_i^2}
	+
	{[\theta_{y,i} - \theta_y(\bfha,t_i)]^2 \over \sigma_i^2}
\right\}.
\end{equation}
The model $\theta_x(\bfha,t_i)$ for the predicted position
is embodied by Eqs. (\ref{abgdef}), (\ref{exact}), (\ref{orbit1}), and
(\ref{nbody}).  The minimization of $\chi^2$ can use standard
algorithms.  We have implemented a slight modification of the
Levenberg-Marquardt method from \citet{Pr88}.  For efficient
location of the minimum, we require a good starting point for the
non-linear search, and estimates of the derivative matrix $d{\bf
\theta}/d\bfa$.

The determination of the derivatives is again greatly simplified for
distant objects.
The gravitational acceleration is, over a $\lesssim10$-year period, a
small perturbation to the inertial motion, barring close encounters.
The inertial motion will be $\sim 2\pi\sqrt\gamma t$ and the
acceleration term will integrate to $\approx4\pi^2\gamma^2t^2/2$.
Furthermore the acceleration is
toward the barycenter, so the transverse component is down by another
factor of $\gamma$.
The angular displacement due to gravity is only a few arcseconds per
year, $\gtrsim300$
times smaller than the inertial motion and reflex motion.
It is clear that in calculating the derivative matrix
$\partial {\bf x} / \partial \bfa$ from Eq.~(\ref{orbit1})
that we can ignore the gravitational term.  All of the derivatives
are thus quite trivial and there is no need to integrate the
derivatives of the orbit.

Selection of a sensible starting point for the minimization is also
simple for distant objects.  We produce a version of the equations of
motion which ignores gravity, is linearized, and also ignores the
$\dot\gamma$ term:
\begin{eqnarray}
\label{linearized}
\theta_x \approx \alpha + \dot\alpha t - \gamma x_E(t), \\
\theta_y \approx  \beta + \dot\beta t - \gamma y_E(t), \nonumber
\end{eqnarray}
These equations are linear in the parameter vector $\bfa$ and
hence have a closed-form minimization requiring only the inversion of
a $5\times5$ matrix \citep{Pr88}.
These very simple equations are actually good to a few arcseconds for
the first year of observation, and in any case provide (with
$\dot\gamma=0$) a good starting point for the non-linear minimization.

Using the analytic derivatives, the Levenberg-Marquardt algorithm also
calculates the covariance matrix ${\bf \Sigma}_\alpha$ of the fitted
parameters.  Since the model $\bft(\bfa,t)$
is dominated by terms linear in the parameters, the linearized
approximation
\begin{equation}
\Delta \chi^2 = (\bfa - \bfha)^T {\bf\Sigma}_\alpha
	(\bfa - \bfha)
\end{equation}
will be accurate even for poorly constrained orbits.  This would not
have been the case had we chosen the orbital-element vector ${\bf
a}$ as the parameter set.

\subsection{Position and Orbital Element Estimation}
Once the best-fit $\bfha$ vector and its covariance matrix
are determined from the fit, determination of orbital elements and
future positions is also simple.  The predicted position at any time
$t$ is $\bft(\bfha,t)$.  The uncertainty ellipse for
the position is specified by the covariance matrix
\begin{equation}
\label{poscovar}
{\bf \Sigma}_\theta = \left( {d\bft(\bfa,t)
	\over d\bfa}\right)^T {\bf \Sigma}_\alpha
	\left( {d\bft(\bfa,t)\over d\bfa}\right).
\end{equation}
The derivatives, as above, are rapidly calculated by ignoring
gravity.  Again the position is nearly linear in our chosen orbital
basis, so this linearized transformation will be accurate even in the
case of large uncertainties in the parameters.

The best-fit orbital element vector ${\bf\hat a}$ is determined from the best-fit
phase-space parameter set $\bfha$ via the usual equations,
which are reviewed in Appendix~\ref{elements}.  The determination of
an uncertainty 
ellipsoid in orbital-element space requires mapping
of the covariance matrix with the derivatives $d{\bf
a}/d\bfa$.  Some of these derivatives are quite complicated
analytic expressions, but can still be coded and
rapidly evaluated.  The map from $\bfa$ to ${\bf a}$
is non-linear, and in some cases degenerate, so the linearized
covariance matrix is formally correct only when the uncertainties in
the elements become small.  We do, however, obtain a useful estimate
of the element uncertainties even for less exact determinations.

\subsubsection{Fitting Singly Degenerate Orbits}
If the arc is too short, the minimization of $\chi^2$ will proceed but
the best-fit parameters may be non-physical.  The degeneracy in the
``acceleration regime'' will be
manifested as one or more large elements of the covariance matrix.  In
particular, several-week orbits will have large values of
$\Sigma_{\dot\gamma \dot\gamma}$.  We recognize these orbits as
unrealistic because they may be unbound.  The assumption of a bound
orbit limits $\dot\gamma$ to ({\it cf.} Eq.~[\ref{energy}])
\begin{equation}
\label{gdotbind}
\dot\gamma^2 \le \dot\gamma^2_{\rm bind} = 
GM_\odot \gamma^3 (1+\gamma^2-2\gamma\cos\beta_0)^{-1/2} -
\dot\alpha^2 - \dot\beta^2.
\end{equation}
A conservative approach to assigning uncertainties in this case is to
assume that
$\dot\gamma$ has a uniform {\it a priori} probability in the
interval $[-\dot\gamma_{\rm bind},+\dot\gamma_{\rm bind}]$.
If, however, we assign
a Gaussian prior distribution to $\dot\gamma$, we can preserve our
least-squares approach to the fitting problem.  We therefore fit by
assuming that $\dot\gamma=0$ with an RMS uncertainty of $\sigma_{\rm
bind}=\dot\gamma_{\rm bind}/\sqrt{3}$.  This value of $\sigma_{\rm
bind}$ has the same variance as the uniform distribution, and the
entire $\pm\dot\gamma_{\rm bind}$ range (plus some marginally unbound
orbits) is contained within the $2\sigma$ contour, so we will obtain a
realistic assessment of the uncertainties.

Our procedure for handling the $\dot\gamma$ degeneracy is thus:
\begin{enumerate}
\item Fit the orbit with all six parameters free to vary.
\item If the unconstrained fit yields $\Sigma_{\dot\gamma
\dot\gamma}<\sigma^2_{\rm bind}$, then $\dot\gamma$ is well
constrained by the observations and we are done.  Otherwise
$\dot\gamma$ is essentially
decoupled from the observations, so we proceed using the binding
constraint instead.
\item Re-fit the observations but set $\dot\gamma=0$ and derive
values and a covariance matrix for the remaining 5 parameters.  Note
that this best-fit orbit is very similar to the V\"ais\"al\"a orbit,
as a line-of-sight velocity near zero implies an orbit near its
perihelion or aphelion.
\label{p5step}
\item Augment the covariance matrix by assigning $\Sigma_{\dot\gamma
\dot\gamma}=\sigma^2_{\rm bind}$, placing zeros in the remaining
off-diagonal elements.
\end{enumerate}
The best-fit parameters $\bfha$ and covariance
${\bf\Sigma_\alpha}$ may then be used to predict positions and orbital
elements as in the non-degenerate case.  The $2\sigma$ error ellipses
will encompass $>95\%$ of all bound orbits that are consistent with
the data, and not be restricted to circular or perihelic orbits.  If
we have some preconceptions about the maximal ellipticity of the
orbits, this may be incorporated into the position predictions by
placing a tighter {\it a priori} constraint on $\dot\gamma$.

\subsubsection{Fitting Doubly Degenerate Orbits}
With only a single or few nights' data, or
if there are only 2 observations, we are in the ``slope regime.'' 
The above technique will fail
due the additional degeneracy between $\dot\alpha$ and $\gamma$ (and
$\dot\beta$ for observations out of the ecliptic plane).  This will
manifest itself in the covariance matrix obtained in Step~\ref{p5step}
above, the 5-parameter fit to the orbit.  The uncertainties
$\Sigma_{\dot\alpha\dot\alpha}$ and/or $\Sigma_{\dot\beta\dot\beta}$
will be large.

In this case we again introduce the binding constraint to limit the
extent of the degeneracy.  We express the transverse kinetic energy
in terms of a parameter $f_b$:
\begin{equation}
\dot\alpha^2 + \dot\beta^2  =  (1+f_b) GM_\odot\gamma^3.
\end{equation}
The binding constraint is then
$-1<f_b<1$, with $f_b=0$ corresponding to a nearly circular orbit.
We crudely implement this constraint by minimizing a new quantity
\begin{equation}
\tilde\chi^2 = \chi^2 + f_b^2 / 3.
\end{equation}
(with $\chi^2$ from Eq. \ref{chisq}),
which pushes the solution toward a circular orbit,
while yielding a covariance matrix that reflects both the
observational uncertainties and the possibility of non-circular
orbits.  More sophisticated constrained optimizations are possible but
there is little point to seeking precise constraints on such a poorly
determined orbit.

\section{Implementation and Examples}
\subsection{C Code}
All of the above algorithms have been implemented in a family of
subroutines and drivers written in C.  These include routines for:
creation and access of binary versions of the DE405 ephemeris
files (provided by David Hoffman); orbit integration; transformations
among the various Cartesian and spherical coordinate systems;
transformations between phase-space and orbital-element bases for
orbits; reduction of positions to topocentric coordinates; and the
fitting procedures described in the previous section.  The code is
portable and self-contained, and is available from the first author
upon request.

The main program is {\tt fit\_radec}, which takes as input a list of
astrometric observations, and executes the orbit-fitting, including
proper handling of degeneracies.  The output is a file specifying the
best-fit $\bfa$ description of the orbit, and its full
covariance matrix ${\bf \Sigma}_\alpha$.

A program {\tt abg\_to\_aei} transforms the $\bfa$ parameters
into traditional orbital elements.  The covariance matrix is also
transformed with a linear approximation.

The $\bfa$ information may be passed to the program {\tt
predict}, which generates predicted positions and their error ellipses
for any specified observation date and site.

The entire process is extremely fast on the typical present-day
desktop computer.  A fit to 275 mock observations of Pluto over a
14-year arc takes about 1 second.  The entire ensemble of observations
of all $\approx300$ KBOs reported to the MPC are fit in a few
seconds.

\subsection{Example and Verification}
\begin{figure}
\plotone{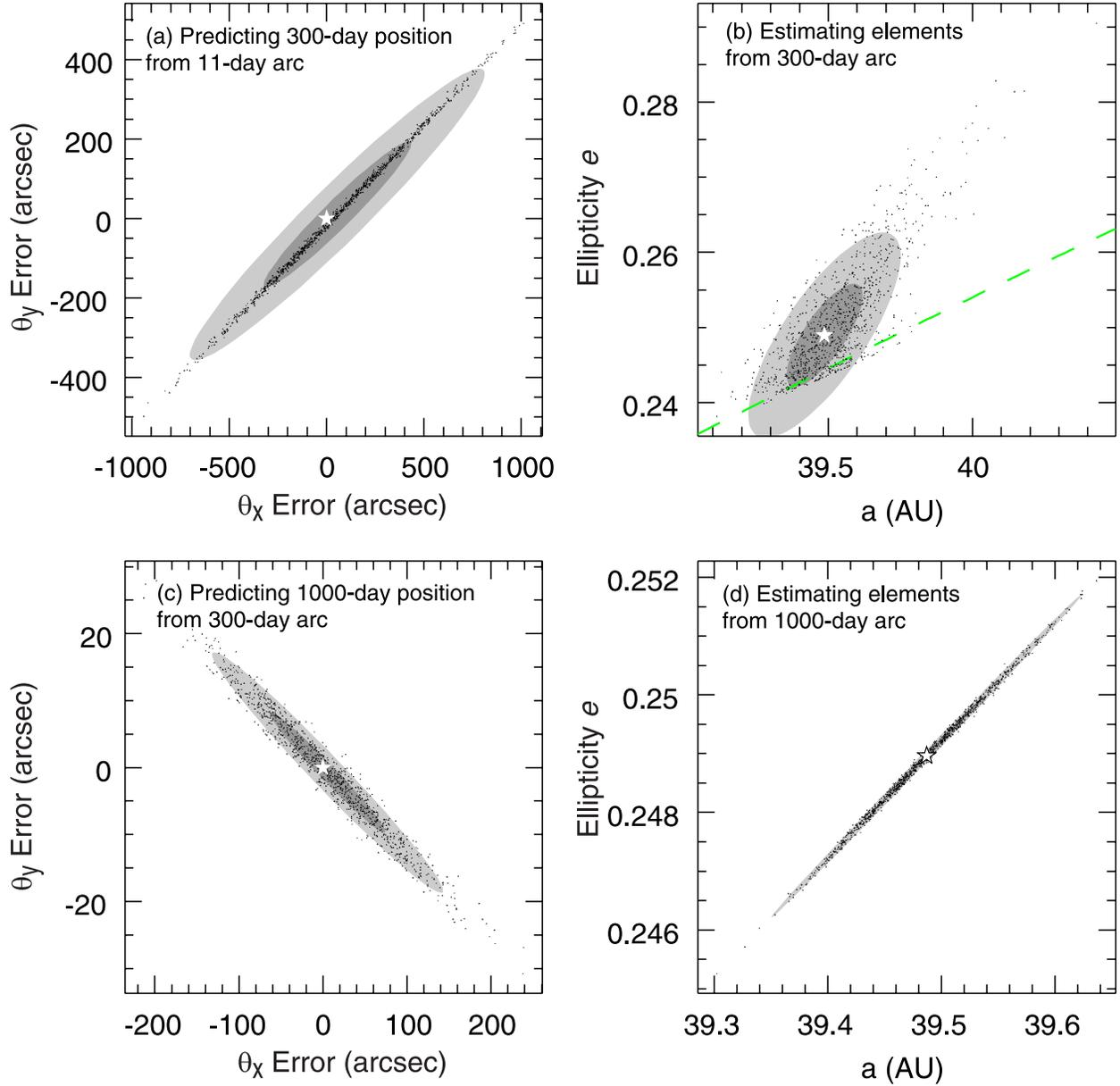}
\caption[dummy]{\small
Demonstration of position/orbital element estimation using the methods
described in the text.  In the left panels (a) and (c), mock
observations of Pluto are used to predict a future position on the
sky.  Each small symbol is the prediction from one mock arc;
the star is the correct position, while the two shaded ellipses are
the 1- and 2-sigma uncertainty regions produced by our methods.  The
panel (b) shows the orbital
parameters $a$ and $e$ estimated from 5 observations over 300 days,
along with the true position and estimated uncertainties.  Panel (d)
shows the elements estimated after an additional observation at 1000
days.  The scatter of estimates matches the uncertainties except in
case (b).  Details of this failure are explained in the text.
}
\label{plutotest}
\end{figure}

We have verified the performance of the algorithms and software by
conducting mock observations of Pluto, Neptune, and several KBOs.
Figure~\ref{plutotest} shows the results of an ensemble of simulated
observations of Pluto from 1995--1998.  The true position of Pluto is
obtained from the DE405 ephemeris; mock observations are created at
specified dates by adding Gaussian deviations to the true positions.
An orbit is then fit to the mock observations.  This fitted orbit is
used to estimate the osculating barycentric elements $a$ and $e$ for
Pluto as well as the uncertainty ellipse in the $a$--$e$ plane.
Alternatively we may use the orbit fit to estimate the
$(\theta_x,\theta_y)$ sky position
of Pluto at some future date and its projected error ellipse.  We can
then examine whether the estimated orbital elements or future position
are consistent with the true ephemeris values.  In all cases we find
the mean $\chi^2$ value of the fit to the orbit
to be as expected from the degrees of freedom
in the fit.

The first panel in Figure~\ref{plutotest} shows the results of using 4
observations over an 11-day arc (JD 2449927--2449938) to predict the
location of Pluto $\approx300$~days after the first observation.  Each point
on the plot is the prediction from one of 1000 realizations of the mock 11-day
arc.  The star shows the true location of Pluto at the 300-day time.
The ellipses show the $1\sigma$ and $2\sigma$ contours of the
estimated error on the position given the observational errors.  The
estimated uncertainty ellipse is quite consistent with the scatter in
the mock predictions.
There appear to be two flaws
in our prediction:  first, the mean of the predicted positions does
not coincide with the true position; second, the minor axis of the
uncertainty ellipse appears to be significantly larger than the scatter.
Both of these phenomena are expected, however, because the 11-day arc
is the ``acceleration regime'' and is degenerate in $\dot\gamma$.
The fits therefore have assumed a value and uncertainty for
$\dot\gamma$.  The displacement of the simulations from the
true value is due to our assumption that $\dot\gamma=0$ for each
fit, which is not the true value.  But the uncertainty ellipse
includes a contribution from our ignorance of $\dot\gamma$,
which makes the ellipse extend beyond the measured scatter, and in
fact properly include the true position.

In the second panel we simulate an observation sequence consisting of
the 11-day arc plus and additional observation at 300 days.  We
plot the estimated $a$ and $e$ values from 1000
mock observations of the 300-day arc.  In this case
the uncertainty ellipses give a roughly correct estimate of the size of the
errors, but the scatter does not match the error ellipses in detail.
This demonstrates a limitation of our method, which is that the 
propagation of the covariance matrix ${\bf \Sigma}_\alpha$ to the
orbital elements is incorrect if the map between them is significantly
nonlinear.  This is the case here, as Pluto is very near perihelion at
this time.  The 300-day arc determines the distance to Pluto quite
well, and hence the scatter plot is bounded by a lower limit to the
perihelion $a(1-e)$.

The third panel shows the results of using the 300-day arc to predict
the position after 1000 days, and the fourth panel shows the scatter
in the $a$--$e$ plane of elements determined from 6 observations on
a 1000-day arc.  We can now usefully constrain all 6 parameters of
the orbit, and the scatter about the true position and orbit are very
accurately described by the estimated covariance matrices.

\section{Optimal Strategies}
\label{optimal}
An optimal schedule for recovery of KBOs would maximize some measure
of the accuracy of the KBO orbit while minimizing the number of
recovery observations required.  A constraint on the observing
schedule is that the
positional uncertainty must not become larger than the field of view
of the instrument, lest the recovery be missed
and the object be lost.  As an illustration of the utility of the
methods described in this paper, we derive an optimal schedule of
observations using a crude algorithm.

We take our measure of the quality of the orbit fit to be the
uncertainty $\sigma_a=\sqrt{\Sigma_{aa}}$ in the semi-major axis of
the orbit.  This is not meant to be a universal metric for quality
but rather a simple example.  We place two constraints on the observing
schedule:  first, observations are not allowed when the solar
elongation of the target is less than 90\arcdeg.  Second, we take the
field of view of the recovery instrument to be $\approx10\arcmin$; then
the 1-sigma positional uncertainty must be $\le2\farcm5$ if the full
2-sigma error ellipse is to be contained within the field of view for
$>95\%$ chance of recovery.  Note that many CCD mosaics
currently on 2-meter and 4-meter class telescopes have fields of view
that substantially exceed 10\arcmin.

A rigourous optimization for a set of $N$ observations would consider
all possible sets of $N$ nights and choose the set which minimizes the
final $\sigma_a$.  We adopt a somewhat simpler
``greedy'' algorithm for optimization as follows:  after observation
$M$, we consider all nights for which observation $M+1$ would be
possible, namely those satisfying the elongation ad positional
uncertainty costraints.  We next calculate the 
$\sigma_a$ which would result from an observation on each possible
night; the $M+1$ observation is then scheduled for the night which
would result in minimal $\sigma_a$.

\begin{figure}
\plotone{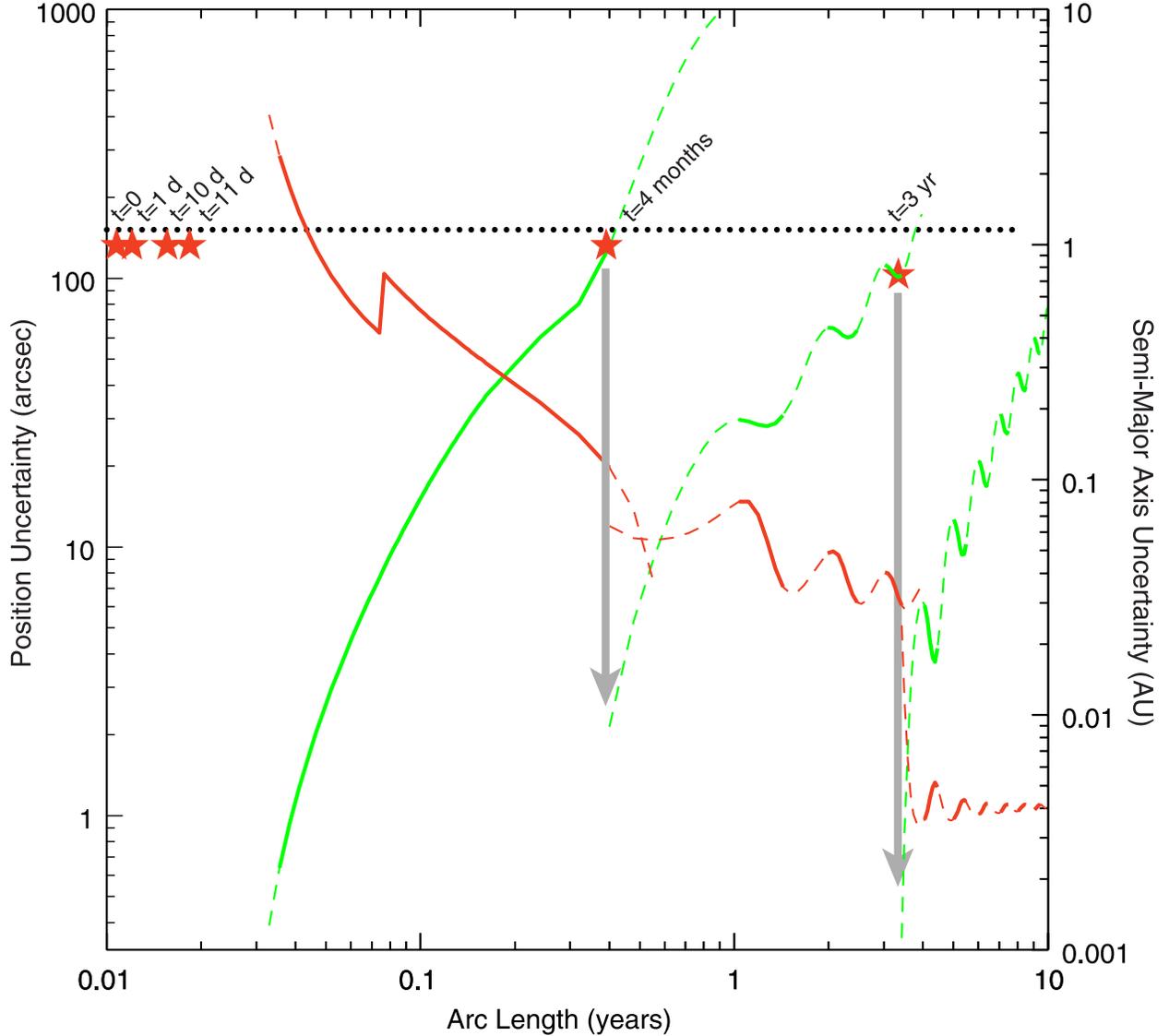}
\caption[dummy]{\small
The observation optimization routine described in the text is applied
here to Pluto.  We assume an initial set of four observations over a 10
days, about 6 weeks before opposition.
We then ask when further observations are best scheduled.  The
upward-sloping lines depict the uncertainty in position of the object
given the 10-day arc; another observation must be done while the
uncertainty is below 150\arcsec\ (the dotted line) in order to assure
successful recovery in a 10\arcmin\ FOV.  The curved is dashed when
the target is too close to the Sun to observe, solid otherwise.  The
downward-sloping lines show the uncertainty in semi-major axis $a$
that would be obtained from a further observation at the given time.
Additional observations (stars) are scheduled at the time during
the recovery window when the measurement will most reduce the
uncertainty on $a$.  Each new
observation then starts a new pair of curves for $a$ and position
uncertainty.  A new observation is needed at 4 months (before the
target disappears behind the Sun), after which the uncertainty on $a$
is only weakly dependent upon the timing of observations.  Another
observation is needed within 3 years else the target becomes lost.
}
\label{plopt}
\end{figure}

Figure~\ref{plopt} illustrates the observation schedule derived for
hypothetical discovery of Pluto in the year 1995.  We assume that the
object is discovered on a pair of nights roughly 45 days before
opposition, then confirmed on another pair of nights 10 days later.
We also presume that each observation has an error of $\sigma=0\farcs2$.
As seen in the Figure, the object must be observed once again in the
discovery
season as its uncertainty is too high to be recovered when it emerges
from behind the Sun the following year.  The object is now
sufficiently well characterized to be found for nearly 3 years.
When is the best time during this window to observe again?  
Interestingly the accuracy on $a$ improves only weakly the longer we
wait for our next observation.  The ``greedy'' strategy suggests we wait 3
years, when the positional uncertainty approaches
2\farcm5.  So after the initial 4 discovery/confirmation images
and only 2 recoveries, $\sigma_a\approx0.03$~AU,
and the orbit is known well enough to permit recovery for the next
decade.  

A counterintuitive result is that the timing of further
observations has little effect upon the accuracy of $a$---only
increased {\it number} and/or {\it accuracy} of observations matters
at this point.  This is because the uncertainty in $a$ derives
primarily from the uncertainty in $\gamma$, {\it i.e.} the distance to
the target.  The leading observable consequence of $\gamma$ 
is the parallax $\gamma x_E(t)$, which does not
grow with time beyond 1 year---hence the very weak improvement with
longer arcs.  Indeed we find that observing near the two quadratures
of the second year yields a $\sigma_a$ that is as low as spacing these
two observations over several years.

Another common observing scheme is to run discovery observations
during two nights near opposition, then attempt confirmation one month
later.  For Pluto, the positional error is sufficiently small this
second month to permit reliable confirmation.  Henceforth the optimal
strategy plays out just as the previous scenario:
a further observation
is required before quadrature to keep the object from being lost while
behind the Sun, and the orbital quality depends primarily upon the
number, not the arc length, of recoveries in later years.

If the astrometric accuracy of the initial 11-day arc can be improved
to 0\farcs1, and the recovery tolerance raised to
$\sigma_a\approx10\arcmin$, then no further observations are required
in the first year.  The object is recoverable at the beginning
of the second season, two observations (near the quadratures) in this
second year reduce $\sigma_a$ to 0.0002~AU, and the position is good
for a decade or more.

These results are of course dependent upon the orbit of the object and
upon the nature of one's metric for orbital quality.  Our optimization
method is, however, easily adapted to other metrics.

\section{Summary}
Orbit fits for KBOs and other distant objects are greatly simplified
by the fact that the gravitational perturbation to the orbit is small
compared both to the inertial motion and to the acceleration of the
Earth.  Gravity can be treated as a perturbation even for decades-long
arcs, which simplifies the fitting process and the propagation of
errors because there is no need to integrate the derivatives of the
orbit.  The orbit itself can be integrated quickly because large time
steps can be taken.  Distant objects move slowly across the sky, so we
can also use the convenience of a tangent-plane projection of the
celestial sphere.  The dominance of reflex motion over proper motion
means that KBO orbit solutions are unambiguous.

Exploiting these properties, we have created algorithms and software
which quickly and accurately calculates orbital elements and
ephemerides {\it and their associated uncertainties} for targets
$\gtrsim10$~AU from the Sun.  In cases where the observed arc is short
enough to leave the orbit degenerate, we can still calculate sensible
positional errors because we have chosen an orbital basis in which the
degeneracy is confined to one parameter, namely the line-of-sight
velocity. For very short arcs in
the ``slope regime,'' 2 additional parameters become degenerate.  
We are still able to place error ellipses on
ephemerides which are conservative in the sense that they will include
any bound orbit.  Most short-arc objects will be recovered closer to the MPC
predicted ephemeris than our error ellipses would suggest, because the
philosophy of the MPC is to use the stronger {\it a priori} consideration
that the orbits resemble those of the majority of the known populations.

Our routines should prove valuable for planning of recovery
observations, as we have illustrated with some simple examples.
Estimates of the uncertainty ellipsoids in orbital-element space will
also be needed for study of the long-term dynamics of the known KBOs.

In the near future we expect that KBO positions will be measured to
accuracies of a few milli-arcseconds with orbiting
observatories---though such observations will be even more expensive
than present-day KBO observations.  It will then be possible to place
strong constraints on KBO orbits even from very short ({\it e.g.}
24-hour) arcs.  For example, the motion of an observatory in low-Earth
orbit imparts a reflex motion on the KBO that is easily detected at
this level, giving a measure of the distance to the target in less
than an hour.  The methods described herein will provide the means to
plan and to analyze such observations.

\acknowledgements
We thank David Hoffman for making available his C code to read the
DE405 ephemerides; Brian Marsden for discussions on fitting of KBO
orbits; Julia Plummer for assistance with Figure~\ref{coords};
and Renu Malhotra and Lynne Allen for serving as ``guinea
pigs'' for tests of the software.
This work is supported by NASA Planetary Astronomy grant
\#NAG5-7860, and grant \#AST-9624592 from the
National Science Foundation.

\newpage

\appendix
\section{Coordinate Transformations}
\label{sphere}
\subsection{Angular Coordinates}
We make use of three sets of coordinates on the celestial sphere:
equatorial coordinates $(\alpha,\delta)$ in the ICRS reference frame,
J2000; ecliptic latitude and longitude $(\ell,b)$, also taken in the
epoch 2000 frame; and our projected coordinates $(\theta_x,\theta_y)$
measured in a tangent plane about some reference direction
$(\alpha_0,\delta_0)$ or $(\ell_0,b_0)$.  The transformation from
equatorial to ecliptic coordinates is implemented as
\begin{eqnarray}
\sin b & = & \cos\epsilon \sin\delta - \sin\epsilon \cos\delta
\sin\alpha \\
\tan\ell & = & {{ \cos\epsilon \cos\delta \sin\alpha +
	\sin\epsilon \sin\delta} \over {\cos\delta \cos\alpha} },
\end{eqnarray}
where $\epsilon=23\fdg43928$ is the obliquity of the ecliptic at J2000
\citep{DE405}. The inverse transformation simply reverses the sign of
$\epsilon$.

The map from ecliptic to tangent-plane coordinates about
$(\ell_0,b_0)$ is
\begin{eqnarray}
\theta_x & = & { {\cos b \sin{(\ell - \ell_0)} } \over
	{\sin b_0 \sin b - \cos b_0 \cos b \cos{(\ell - \ell_0)} } }
	\\
\theta_y & = & {{ \cos b_0 \cos b - \sin b_0 \sin b \cos{(\ell -
	\ell_0)} } \over
	{\sin b_0 \sin b + \cos b_0 \cos b \cos{(\ell - \ell_0)} } }.
\end{eqnarray}
The inverse transformation is
\begin{eqnarray}
\sin b & = & { { \sin b_0 + \theta_y \cos b_0} \over
	\sqrt{1+\theta_x^2+\theta_y^2} } \\
\sin {(\ell - \ell_0)} & = & {{ \theta_x / \cos b_0} \over
	\sqrt{1+\theta_x^2+\theta_y^2}}  .
\end{eqnarray}
Partial derivative matrices are straightforward for all
transformations, and approximations for $\theta_x, \theta_y \ll 1$ are
particularly simple.

\subsection{Spatial Coordinates}
There are also three relevant spatial coordinate systems.  The
equatorial ${\bf x}_{\rm eq}$ has origin at the Solar System
barycenter, with $x$ positive to the Vernal Equinox and $z$ positive
to the North equatorial pole of J2000.  The ecliptic system ${\bf
x}_{\rm ec}$ also has origin at the barycenter, but $z$ is positive to
the North ecliptic pole of J2000.  Our ``telescope-centric''
coordinate system ${\bf x}_T$ 
has origin at the location of the observer at $t=0$,
with $z$ positive along the line of sight to the target at $t=0$, the
$y$ axis in the plane of $z$ and the North ecliptic pole---see
Figure~\ref{coords}. 

The transformation from equatorial to ecliptic coordinates is

\begin{equation} 
\left[ 
\begin{array}{c} 
x \\ 
y \\ 
z \\ 
\end{array} 
\right]_{\rm ec} 
= 
\left[ 
\begin{array}{ccc} 
1 & 0 & 0 \\ 
0 & \cos\epsilon & \sin\epsilon \\ 
0 & -\sin\epsilon & \cos\epsilon \\
\end{array} 
\right]   
\left[ 
\begin{array}{c} 
x \\ 
y \\ 
z \\ 
\end{array}  
\right]_{\rm eq} 
\end{equation}
The inverse transformation inverts the sign of $\epsilon$.

The transformation from ecliptic to telescope-centric coordinates is 
\begin{equation} 
\left[ 
\begin{array}{c} 
x \\ 
y \\ 
z \\ 
\end{array} 
\right]_T
= 
\left[ 
\begin{array}{ccc} 
-\sin\ell_0 & \cos\ell_0 & 0 \\ 
-\cos\ell_0 \sin b_0 & -\sin\ell_0 \sin b_0 & \cos\ell_0 \\
\cos\ell_0 \cos b_0 & \sin\ell_0 \cos b_0 & \sin\ell_0 \\
\end{array} 
\right]   
\left[ 
\begin{array}{c} 
x-x_0 \\ 
y-y_0 \\ 
z-z_0 \\ 
\end{array}  
\right]_{\rm ec}.
\end{equation}
The ecliptic-coordinate location of the observer at time $t=0$ is
given by the vector $\{x_0,y_0,z_0\}$.  The inverse transformation is
\begin{equation} 
\label{Ttoec}
\left[ 
\begin{array}{c} 
x-x_0 \\ 
y-y_0 \\ 
z-z_0 \\ 
\end{array}  
\right]_{\rm ec}
= 
\left[ 
\begin{array}{ccc} 
-\sin\ell_0 & -\cos\ell_0 \sin b_0 & \cos\ell_0 \cos b_0 \\ 
\cos\ell_0 & -\sin\ell_0 \sin b_0 & \sin\ell_0 \cos b_0\\
0 & \cos b_0 & \sin b_0 \\
\end{array} 
\right]   
\left[ 
\begin{array}{c} 
x \\ 
y \\ 
z \\ 
\end{array} 
\right]_T.
\end{equation}
Since all the transformations are linear, the partial derivatives are
trivial. 

\section{Orbit Basis Transformations}
\label{elements}
The orbit-fitting algorithms produce a description of the orbit in
terms of $\bfa
=\{\alpha,\beta,\gamma,\dot\alpha, \dot\beta, \dot\gamma\}$.  To
obtain orbital elements, we first transform to Cartesian phase space
elements in the telescope-centric coordinate system:
\begin{equation}
\begin{array}{lll}
x_T = \alpha / \gamma & y_T = \beta / \gamma & z_T = 1 / \gamma \\
\dot x_T = \dot\alpha / \gamma & \dot y_T = \dot\beta / \gamma & 
\dot z_T = \dot\gamma / \gamma \\
\end{array}
\end{equation}
The telescope-centric phase space coordinates can then be transformed
to barycentric
ecliptic Cartesian phase-space coordiates using Eq.~(\ref{Ttoec}).
The transformation matrix for the velocities is the same as that for
the position, but there is no telescope-to-barycenter offset for the
velocities.

With the barycentric position and
velocity ${\bf x}$ and ${\bf v}$
in hand, the osculating orbital elements are determined with the
standard formulae below.  Using the gravitational constant
$\mu\equiv GM$, we have

Semimajor axis:
\begin{equation} 
a^{-1}={2 \over {x}} - { {v^2} \over \mu }.
\end{equation}
Eccentricity:
\begin{eqnarray}
{\bf e} & \equiv &
 \left[ {{v^2} \over \mu}
	- { 1 \over {x}} \right] {\bf x} -
	 \left[ {{\bf x} \cdot {\bf v} \over  \mu } \right] {\bf v}\\
e & = & |{\bf e}|
\end{eqnarray}

With the further definitions
\begin{eqnarray}
{\bf h} & \equiv & {\bf x} \times {\bf v} \\
{\bf n} & \equiv & {\bf \hat z} \times {\bf h}
\end{eqnarray}
the inclination, ascending node, and argument of perihelion are given by
\begin{eqnarray}
\cos i & = & { {h_z} \over  h } \\
\cos \Omega & = & { {n_x} \over  n } \\
\cos \omega & = & { { {\bf n}\cdot{\bf e}} \over  {n e} }.
\end{eqnarray}
The eccentric and mean anomalies are
\begin{eqnarray}
\tan E & = & \frac{\bar{y}/b}{(\bar{x}/a)+e} \\ 
M & = & E-e\sin E,
\end{eqnarray}
with
\begin{eqnarray}
\bar{x} & = & \frac{p-x}{e} \\
\bar{y} & = & \frac{{\bf x}\cdot{\bf v}}{e}\sqrt{\frac{p}{\mu}} \\ 
b & = & a\sqrt{1-e^2} \\ 
p & = & \frac{h^2}{\mu}  
\end{eqnarray} 
Finally, the time of periapse passage is
\begin{equation}
T_p  =  t_0- M\sqrt{a^3/\mu},
\end{equation}
where $t_0$ is the time at which ${\bf x}$ and ${\bf v}$ are determined.

The partial derivatives
necessary for propagation of errors are 
tedious but calculable.  Our phase-space elements are centered on the
solar system barycenter, therefore the osculating elements are also
barycentric. We take $M$ as total mass of the solar system
since we are using the barycenter.

\end{document}